\documentclass[]{acmart}
\AtBeginDocument{%
  }

\copyrightyear{2026}
\acmYear{2026}
\setcopyright{cc}
\setcctype{by-nc-nd}
\acmConference[IDC '26]{Proceedings of the 25th Interaction Design and Children Conference}{June 22--25, 2026}{Brighton, United Kingdom}
\acmBooktitle{Proceedings of the 25th Interaction Design and Children Conference (IDC '26), June 22--25, 2026, Brighton, United Kingdom}
\acmDOI{10.1145/3773077.3806107}
\acmISBN{979-8-4007-2283-7/2026/06}
\acmISBN{978-1-4503-XXXX-X/2018/06}
\usepackage{booktabs}
\usepackage{multirow}

\begin{document}

\title[Building to Understand]{Building to Understand: Examining Teens' Technical and Socio-Ethical Pieces of Understandings in the Construction of Small Generative Language Models}

\author{Luis Morales-Navarro}
\orcid{0000-0002-8777-2374}
\email{luismn@upenn.edu}
\affiliation{%
  \institution{University of Pennsylvania}
  \city{Philadelphia}
  \state{Pennsylvania}
  \country{USA}
}

\author{Daniel J. Noh}
\orcid{0009-0002-7219-1988}
\email{dnoh@upenn.edu}
\affiliation{%
  \institution{University of Pennsylvania}
  \city{Philadelphia}
  \state{Pennsylvania}
  \country{USA}
}

\author{Lucianne Servat}
\orcid{0009-0005-7842-0149}
\email{lucianne@upenn.edu}
\affiliation{%
  \institution{University of Pennsylvania}
  \city{Philadelphia}
  \state{Pennsylvania}
  \country{USA}
}

\author{Carly Netting}
\orcid{0009-0005-1438-9031}
\email{cnetting@fi.edu}
\affiliation{%
  \institution{The Franklin Institute}
  \city{Philadelphia}
  \state{Pennsylvania}
  \country{USA}
}

\author{Yasmin B. Kafai}
\orcid{0000-0003-4018-0491}
\email{kafai@upenn.edu}
\affiliation{%
  \institution{University of Pennsylvania}
  \city{Philadelphia}
  \state{Pennsylvania}
  \country{USA}
}

\author{Dana{\'e} Metaxa}
\orcid{0000-0001-9359-6090}
\email{metaxa@upenn.edu}
\affiliation{%
  \institution{University of Pennsylvania}
  \city{Philadelphia}
  \state{Pennsylvania}
  \country{USA}
}

\renewcommand{\shortauthors}{Morales-Navarro et al.}

\begin{abstract}
The rising adoption of generative AI/ML technologies increases the need to support teens in developing AI/ML literacies. Child-computer interaction research argues that construction activities can support young people in understanding these systems and their implications. Recent exploratory studies demonstrate the feasibility of engaging teens in the construction of very small generative language models (LMs). However, it is unclear how constructing such models may foster the development of teens' understanding of these systems from technical and socio-ethical perspectives. We conducted a week-long participatory design workshop in which sixteen teenagers constructed very small LMs to generate recipes, screenplays, and songs. Using thematic analysis, we identified technical and socio-ethical pieces of understandings that teens exhibited while designing generative LMs. This paper contributes (a) evidence of the kinds of pieces of understandings that teens have when constructing LMs and (b) a theory-backed framing to study novices' understandings of AI/ML systems.
\end{abstract}

\ccsdesc[500]{Human-centered computing~Empirical studies in HCI}
\ccsdesc[300]{Social and professional topics~Computing literacy}
\ccsdesc[300]{Social and professional topics~K-12 education}

\keywords{AI literacy, language models, teens, LLMs, computational empowerment, understanding, sensemaking, ethics, knowledge-in-pieces, artificial intelligence, machine learning}

\maketitle

\section{Introduction}

The widespread adoption of generative AI/ML technologies has increased the urgent need for supporting young people in developing AI/ML literacies \citep{long2020ai, touretzky2023machine}. Recent frameworks highlight that AI/ML literacies involve four domains: engaging with (recognizing, using and evaluating the outcomes of AI/ML systems), creating with (using AI/ML in creative and problem-solving processes), managing (deciding and overseeing which tasks should be assigned to AI systems), and designing AI/ML (designing AI/ML systems to understand how these work) \cite{oecd2025empowering}. Yet, much of the emphasis in AI/ML literacy efforts has centered on preparing young people to \textit{use} these systems, with less attention to their participation in designing or building them. For example, research conducted on teens' understanding of generative language models (LMs), such as ChatGPT, has commonly focused on their everyday interactions as \textit{users} of systems \citep{solyst2024children,marx2024identifying}. However, within child-computer interaction (CCI), computational empowerment researchers argue that engaging young people in design activities is critical, as learners can develop their agency and build a deeper understanding of sociotechnical systems through the construction of artifacts \citep{dindler2020computational}. Here, designing AI/ML involves understanding the design decisions that shape model behaviors, that systems make predictions based on data, collecting and curating data to train models, evaluating the outputs of a system, and considering the systems' potential impact and limitations \cite{oecd2025empowering}. Recent exploratory studies have demonstrated that it is indeed feasible to engage teens in the construction of very small generative LMs~\cite{MORALESNAVARRO2025100769, 10.1145/3769994.3770061}. Considering the exploratory nature of prior studies, however, more research is needed to determine how such construction activities, in which teens take on the role of \textit{designers}, may foster or support teens' understanding of LMs. Thus, this paper presents an investigation of teenagers' engagement with the construction of very small LMs, focusing on how designing these systems supports their understandings of such systems.

Additionally, studies of teens' understanding of LMs from the perspective of \textit{users} have investigated the technical and functional understandings young people have of how models work \citep{marx2024identifying} and their socio-ethical stances\footnote{Those related to teens' understandings of ethics, values, social norms and the social implications of AI/ML systems \cite{10.1145/3708550.3730560, 10.1145/3769994.3770037}.} towards these systems \citep{solyst2024children} separately. However, technical and socio-ethical issues of machine learning are closely intertwined \citep{10.1145/3531146.3533158}. As such, there is a critical need to explore how learners’ technical and socio-ethical understandings intersect. To do so, we propose an in-pieces approach to study novices' understandings of AI/ML systems. Specifically, in this paper, we build on prior research on knowledge-in-pieces \citep{disessa2004coherence}, ideology-in-pieces \citep{philip2011ideology}, and folk theories of sociotechnical systems \citep{10.1145/3173574.3173694} to propose an in-pieces approach to study technical and socio-ethical understandings in tandem. An in-pieces framing can reveal how learners draw on diverse and fragmented ideas when reasoning about AI/ML systems and how technical and socio-ethical pieces of understandings emerge unevenly in practice. 

We conducted a participatory design workshop with sixteen 14-15-year-old teens enrolled in a four-year after-school program at a science center in the Northeastern United States. During the workshop, we took a data-driven approach to AI/ML education that glassboxes how data shapes model performance and blackboxes the role of learning algorithms in the ML pipeline \citep{10.1145/3459990.3460712, olari2024data}. Teens iteratively built small datasets (22-230K tokens) to train very small LMs using the nanoGPT framework \citep{KaparthyNano}. Using thematic analysis, we address the following research question: \textbf{What socio-ethical and technical pieces of understanding do teens exhibit in-the-moment as they design generative LMs?} Our analysis shows that socio-ethical and technical pieces of understanding were often connected to each other, sometimes divergent, and highly influenced by contextual aspects.  
This paper contributes (a) evidence of the kinds of understandings that teens have when constructing very small generative LMs; (b) a theory-backed framing to study novices' fragmented, unstable understanding of AI/ML systems; and (c) directions for future designs and research to foster teens' understanding of LMs.  

\section{Background}
Here, we situate our work within efforts in CCI that engage young people in the construction of generative LMs, highlighting the need for evidence of the kinds of understandings that young people have when they participate in such construction activities. Following, we frame model building as a sociotechnical activity that requires both technical and socio-ethical understandings. Then, we review how prior research has approached young people's understandings of AI/ML. And finally, we propose an in-pieces approach that can be used to study both technical and socio-ethical understandings of LMs. 

\subsection{Engaging Young People in the Construction of Generative Language Models}
CCI research on computational empowerment (CE) emphasizes that learning about computing should go beyond understanding the technical mechanisms, shifting the focus from solely learning technical skills to providing learners opportunities to participate in the development of computing systems that are relevant to their everyday lives \cite{dindler2020computational, iversen2018computational}. CE centers on construction and deconstruction participatory design activities in which learners are ``protagonists'' \cite{smith2023research}, or where learners are the ``main agents'' in designing and evaluating systems \cite{iversen2017child}. In this paper, we focus on the construction (that is, the design of) small generative LMs. 
CE researchers conjecture that through construction activities, learners develop deeper understandings of technology and its broader implications~\cite{iversen2017child}. Traditionally, it is assumed that construction activities are limited to learning about technical aspects of designing computing systems. However, as \citet{iivari2023computational} demonstrated, construction also involves critically analyzing design ideas and considering socio-ethical implications. 

Involving young people as designers of AI/ML systems has a long history going back to symbolic AI research in the 1970s, but it is only over the past decade that studies in CCI and computing education have investigated how young people can be engaged in the design of ML models through a data-driven approach \cite{kahn2021constructionism, druga2021children}. This approach glassboxes, or makes transparent, how data shapes model performance by having learners build datasets to train models and blackboxes the role of learning algorithms in the AI/ML pipeline \cite{vartiainen2021machine}.  In recent studies with data-driven approaches, participants create small datasets that can be easily and quickly refined to train models and improve performance \cite{zimmermann-niefield_youth_2020, tseng2024co, bilstrup2024ml}. Such research has also been extended to natural language processing, in which teens label semantic data and train classifiers \cite{druga2018growing, hjorth2021naturallanguageprocesing4all, norouzi_lessons_2020, katuka2024integrating}. To our knowledge, little research has been conducted on how teens and children engage in building generative LMs.

In the past year, researchers have turned their attention to small LMs, running exploratory studies and creating tool prototypes to engage learners in the construction of generative models \cite{MORALESNAVARRO2025100769, 10.1145/3737609.3747112, 10.1145/3713043.3728857, 10.1145/3769994.3770061}. For example, \citet{10.1145/3737609.3747112} introduced the Engine Room, a collection of paper- and web-based tools for learners to create LMs using Markov chains. These tools were successfully integrated into language arts courses where teachers scaffolded students to discuss and reflect on the differences between human writing and generative LMs \cite{10.1145/3713043.3728857}. In a different study, \citet{MORALESNAVARRO2025100769} conducted participatory design sessions with teenagers in which they designed babyGPTs, or very small LMs, using the nanoGPT framework to generate screenplays. Through the design process, participants engaged with key AI/ML data practices (quality control, preparing data, implementing solutions) and ethical issues (copyright, environmental impact, misinformation). \citet{10.1145/3769994.3770061} proposed a teachable machine for LMs that enables novices to train small LMs in the browser. While these exploratory studies and tool prototypes highlight the feasibility of engaging young learners in designing generative LMs, there is a need for further research on what learners understand about generative language from both technical and socio-ethical perspectives when participating in design activities. 

\subsection{Model Building as a Sociotechnical Activity}
Many AI/ML literacies frameworks emphasize the importance of engaging young people in thinking about ethical and societal issues \cite{touretzky2019envisioning, oecd2025empowering}, yet such efforts are often disconnected from model-building activities \cite{10.1145/3702652.3744217}. A few exceptions include studies in which young people evaluate each other's projects and models \cite{kahila2024pedagogical
}, create ethical matrices for their own AI/ML projects \cite{williams2023ai+}, or use cards with ethical questions to reflect on their own projects \cite{bilstrup2020staging}. Taking a sociotechnical approach to model building is imperative, as potentially harmful societal consequences and implications of AI/ML systems are closely intertwined with functionality failures \citep{fiesler2020ethical, petrozzino2021pays, 10.1145/3531146.3533158}. However, embedding socio-ethical concerns in the model-building process presents challenges even for experts, as ethics and societal considerations are often perceived as personal endeavors or as obstacles that hinder progress in organizational cultures that prioritize efficiency, speed, and innovation \cite{deng2025supporting, avnoon2024contextualizing, 10.1145/3593013.3593990}. Additionally, model designers often lack the necessary infrastructure and support to deeply engage with the socio-ethical implications of their own projects \cite{deng2025supporting}. As such, `critical awakenings' and interest in socio-ethical issues are seen as individual efforts to combat the pervasive techno-optimism and techno-solutionism of the AI/ML industry \cite{malik2022critical}.  Such difficulties are also present in computing education with young people, who are often introduced to computing and model building as value-neutral and strictly technical activities \cite{veldhuis2025critical, ko2021critically, rucker2026taking}. Addressing these challenges not only involves designing interventions that support socio-ethical thinking when learners build models but also having frameworks that support the study of novices' understandings of AI/ML from both socio-ethical and technical perspectives. In the following sections, we review how young people's understandings of AI/ML systems have been studied in prior research and propose taking an in-pieces approach that can support the analysis of both socio-ethical and technical understandings. 

\subsection{Young People's Understanding of AI/ML}

With the widespread adoption of AI/ML systems, researchers have started to investigate how young people make sense of AI/ML by looking into their everyday understandings \cite{marx2024identifying, muhling2023novices}. In this section, we review three ways in which young people's understanding of AI/ML is framed in existing research: attributions, intuitive theories and misconceptions, and critical conceptions.

\subsubsection{Attributions framing}
Several studies have investigated young people's understandings of AI/ML systems by examining how they attribute human and non-human characteristics to AI/ML applications. Studies from the attribution perspective center on young people's conceptions of the anthropomorphic qualities of AI/ML systems, such as sentience, emotion, abilities, and intelligence. As such, these studies often investigate children's perception of AI/ML systems in terms of intelligence or ``smartness'' \cite{druga2018smart, andries2023alexa, williams2019popbots, williams2019artificial} or in comparison to human abilities \cite{festerling2020alexa, van2021alexa}. While these studies provide insights into young people's perceptions of AI/ML, they do not account for understandings of systems' inner workings.

\subsubsection{Functional framing}
Far less attention has been given to young people's everyday understandings of how AI/ML systems work, with most of this research investigating how young people build intuitive theories. From these theories, researchers derive misconceptions that arise when the theories do not align with the scientifically-accepted explanations. 
 
\paragraph{Intuitive Theories} Researchers have studied how young people build theories through their everyday interactions with AI/ML systems. This work highlights that young people use pre-existing knowledge to build theories or hypotheses about AI/ML systems, including egocentric hypotheses (projecting their own thinking processes and actions on AI/ML systems) and observational ad hoc hypotheses (based on observed behaviors of AI/ML systems) \cite{druga2021children}. Other theories involve ideas about data storage and access \cite{szczuka2022children}, and decision-making processes based on data \cite{lin2025children}. More recently, \citet{10.1145/3713043.3728856} described three common mental models that children have about how AI/ML systems reason: inductive reasoning (AI/ML systems generalize patterns from data to make predictions), deductive reasoning (AI/ML systems apply predefined rules), and inherent reasoning (AI/ML systems are inherently intelligent). 

While these studies contribute detailed evidence on children’s understanding of AI/ML, they align with research in conceptual development, which argues that children organize everyday understandings to build ``a cohesive, unitary theory that might contain misconceptions of scientific information.'' \cite{vosniadou2019development} Such an approach has led to the idea that intuitive theories that differ from scientific explanations might make it harder for learners to adopt formal understandings \cite{clement1982students, mccloskey2014naive, hitron2019can}.

\paragraph{Misconceptions}
Over the past five years, a number of studies have investigated young people’s misconceptions of AI/ML systems. Researchers argue that interacting with black-boxed AI/ML systems may lead young people to develop inaccurate or simplified mental models that can be difficult to overcome in instruction \cite{hitron2019can, long2020ai}. While enumerating all the misconceptions present in the literature is beyond the scope of this paper, we review some common misconceptions and invite readers to follow up on the references. 

Several studies highlight that learners often assume that AI/ML systems learn like humans and possess intuition and common sense \cite{marx2024identifying, mertala2022finnish}. A variation of this misconception is what \citet{marx2024identifying} call ``continuous learning,'' which involves the idea that AI/ML systems learn autonomously and improve themselves while being used. Previous research, however, argues that novices may also hold the opposite misconception: that humans decide on all the specifications and program all possible behaviors of AI/ML systems \cite{mertala2024finnish, lin2025children, marx2024identifying}. Other misconceptions include arguing that AI/ML systems always provide exact, deterministic outputs \cite{marx2024identifying}. Some studies highlight that learners may assume that the raw data used to train an AI/ML system is stored and searched by the system every time the system is used \cite{marx2024identifying}.

While misconceptions have received increasing research attention in AI/ML education and child-computer interaction, research in the learning sciences has long questioned the deficit orientation of viewing novices and young people as holders of flawed ideas \cite{pea1986language, smith1994misconceptions, Linn_2005, clark2013knowledge}. For instance, \citet{pea1986language} explains that ``it is not that students literally believe that the computer has a mind'' but that this is an analogy that is expected, as novices do not have exposure to technical knowledge and language to be able to build more technically accurate explanations. 

\subsubsection{Critical framing}
Recent studies have investigated young people's everyday critical and socio-ethical understandings. Several studies have investigated youth’s perceptions of fairness and potentially harmful algorithmic biases \cite{salac2023scaffolding, Salac2023Funds, lee2022eliciting, coenraad2022s, 10.1145/3762807, heeg2025young}. Other researchers have explored broad socio-ethical stances that youth may have about AI/ML systems, related to humans becoming over-dependent on AI/ML systems\cite{solyst2023potential}, concerns about privacy \cite{ko2024youth, lin2025children, heeg2025young}, misinformation and fact-checking \cite{10.1145/3762807}, mechanisms for human oversight \cite{10.1145/3762807}, and the regulation of AI/ML systems \cite{ko2024youth}. Whereas this work shows diverse ways in which young people understand critical and socio-ethical aspects of AI/ML systems, none of the studies cited above addressed the relationship between these critical and technical understandings that youth may have. Here, we see the need for considering socio-ethical understandings as equally important to technical understandings \cite{rucker2023modeling} and the potential for an integrative approach that can enable us to study both socio-ethical and technical understandings of AI/ML.

\subsection{An In-Pieces Approach}
We propose taking an in-pieces approach to novice understandings that accounts for incoherence and context, recognizes the value of learners' existing understandings, centers on how systems work, and can be used to study both technical and socio-ethical understandings in tandem. While current perspectives offer insights into how young people perceive AI/ML, the focus on attributions, intuitive theories, and misconceptions may also restrict our understanding of how young people's everyday understandings can be leveraged for the design of learning tools and activities. In this section we bring together research on fragmented understanding of scientific phenomena \cite{disessa2004coherence} with research on fragmented socio-ethical and ideological thinking \cite{philip2011ideology} and on people's fragmented understanding of socio-technical systems \cite{10.1145/3173574.3173694} to argue that studying teens' technical and socio-ethical understanding of language through an in-pieces approach can be promising to understand how such understandings are constructed, relate to each other, and emerge in context.

In his knowledge-in-pieces work, \citet{disessa1993toward} argued that everyday intuitive knowledge could be organized as a collection of pieces of knowledge or phenomenological primitives (p-prims). These pieces enable people to explain and predict scientific phenomena intuitively and in a way that is consistent with their lived experiences. Yet, these exist in a complex system and are contextually activated; as such, novices may have inconsistent understandings developed across different contexts. Knowledge pieces are part of a ``rich naïve cognitive ecology [that] constitutes a generative pool of resources'' (p. 43) that can be productive for learners to attain formal scientific knowledge. Here, learners develop a repertoire of pieces of knowledge, adding understandings from instruction to those from their lived experiences while grappling with multiple and sometimes conflicting conceptions they may have about scientific phenomena \citep{Linn_2005}.

Recently, researchers have used \citeauthor{disessa2004coherence}'s \cite{disessa2004coherence} in-pieces theory to study teens’ understanding of AI/ML systems \cite{rosenbaum2024knowledge}. For instance, Chao and Jiang \cite{rosenbaum2024knowledge} have investigated how teens make sense of text classification models, noting that learners often use their conceptions of words and their meanings to make sense of the models; this coupling, they argue, may be an important knowledge piece that should be considered in natural language processing education. \citet{morales2024investigating} investigated teens' everyday conceptions of AI/ML systems, such as TikTok filters and voice assistants, from an in-pieces perspective. Their analyses revealed that teens showed some understanding that ML systems learn from training data and that when being used, applications recognize patterns in input data and, depending on these, provide different outputs. We build on this work by extending it to consider socio-ethical understandings; to do so, we take into account research on ideology-in-pieces and algorithmic folk theories that have been used to study people's socio-ethical thinking. 

\citet{philip2011ideology} synthesized diSessa's work with theories of ideology, proposing the framework of ``ideology-in-pieces'' to study people's sensemaking of socio-ethical issues. He argues that, unlike traditional concepts in science, ideas like justice, ethics, blame, and responsibility do not have normative meanings---rather, their meanings are constructed by communities. Philip claims that, when making sense of critical issues related to justice, for example, people rely on existing conceptions that are fragmented, socially situated within particular systems of power, and activated in specific contexts. Therefore, they are not always applied in consistent manners, with people exhibiting different socio-ethical understandings in different contexts: ``Learning to use a concept in a particular context entails recognizing relevant features and making meaningful inferences in that context'' \citep{philip2011ideology}. This framing of socio-ethical thinking can be productive for the study of socio-ethical conceptions that learners may have about ML systems.

\citeauthor{10.1145/3173574.3173694}'s \cite{10.1145/3173574.3173694} conceptualization of algorithmic folk theories is based on the idea that people have fragmented understandings of sociotechnical systems. They argue that people's understandings have a ``fragmentary nature'' characterized by instability and incoherence, where ``pieces of information'' activated together are combined to build folk theories. \citet{muhling2023novices} addressed some of the contradictions and incoherence that may emerge in everyday understandings of ML systems, explaining that students draw on in-this-particular-context ideas to explain the phenomena they observe. Specifically, they built on earlier work by \citet{rucker2016review} that recognizes that ``an individual may hold several conceptions simultaneously, which may or may not be selected in a given context or situation.'' 

Building on these theories, we argue that it is possible to study teens' understanding of LMs by identifying pieces of technical and socio-ethical understandings and observing how these pieces relate to each other and are exhibited in the context of building models. As such, we ask: \textbf{What socio-ethical and technical pieces of understanding do teens exhibit in-the-moment as they design generative small LMs?} 
\section{Methods}
Here, we describe the context of our research, the participatory design activities that teens engaged with, and our data collection and analysis approach.

\subsection{Context and Participants}
We conducted participatory design activities with teens at a Northeastern US science center. These included four two-hour workshops in March and April 2025 and a two-week (4-hour-a-day) intensive summer workshop in July 2025. The science center educators invited the teens to take part in the study. Nineteen of the twenty-two fourteen-to-fifteen-year-olds enrolled in the program gave parental consent and assented to participate in research, and three did not attend the summer workshop.  As such, this study involved sixteen teenagers (n=16). 

After the workshop, participants completed a demographic worksheet with questions about their prior computing and AI/ML experiences, use of LLMs, race, gender, and age (Table \ref{tab:demo}). All questions included a fill-in-the-blank option, and teens were given the choice to abstain from answering any of them. In terms of race and ethnicity, nine teens identified as Black/African-American, four as Asian/Pacific Islander, three as Latinx/Hispanic, and two as White, with two choosing more than one category. Eleven teens identified as men, and five identified as women. All teens reported using LLM-powered systems, with four using them everyday, two twice a week, five once a week, two every other week, and three once a month. Eleven teens reported that they had taken prior computer science courses or workshops, four had taken prior AI/ML courses or workshops, and nine had learned something about AI/ML in one of their classes at school.\footnote{E.g., A language arts teacher talked about ChatGPT in one lesson.} Finally, we asked all teens to decide on their own pseudonyms \cite{kivuva2024cultural}---fourteen teens provided their own, and two asked researchers to decide. The Institutional Review Board of the University of Pennsylvania approved the study protocol.

\begin{table}[]
\caption{Demographic Information}
\label{tab:demo}
\begin{tabular}{lrllll}
\hline
\textbf{Pseudonym}      & \multicolumn{1}{l}{\textbf{Age}}     & \textbf{Gender}     & \textbf{Race}                                                                          & \textbf{\begin{tabular}[c]{@{}l@{}}Use of LLM-powered\\ Systems\end{tabular}}     & \textbf{\begin{tabular}[c]{@{}l@{}}Prior \\ Experiences\end{tabular}}     \\ \hline
Brianna                 & 15                                   & Woman               & \begin{tabular}[c]{@{}l@{}}Asian/Pacific Islander, \\ White\end{tabular}               & Once a month                                                                      & CS, AIC                                                                   \\
Butterfly               & 14                                   & Woman               & Black/African American                                                                 & Every other week                                                                  & AIC                                                                       \\
Darwin                  & 15                                   & Man                 & Hispanic/Latinx                                                                        & Once a week                                                                       & AIC                                                                       \\
Drummer Boy             & 15                                   & Man                 & Hispanic/Latinx                                                                        & Once a month                                                                      &                                                                           \\
Dunkin Lover            & 15                                   & Woman               & Black/African American                                                                 & Everyday                                                                          & CS, AI/ML, AIC                                                            \\
Gregory                 & 15                                   & Man                 & Black/African American                                                                 & Twice a week                                                                      & CS, AIC                                                                   \\
India                   & 15                                   & Woman               & Black/African American                                                                 & Twice a week                                                                      & CS                                                                        \\
Jason                   & 15                                   & Man                 & Black/African American                                                                 & Once a month                                                                      & CS                                                                        \\
Jimmy Bob               & 15                                   & Man                 & White                                                                                  & Once a week                                                                       & CS, AIC                                                                   \\
King                    & 15                                   & Man                 & Black/African American                                                                 & Everyday                                                                          & CS, AI/ML, AIC                                                            \\
Logan                   & 14                                   & Man                 & Black/African American                                                                 & Once a week                                                                       & CS, AIC                                                                   \\
Meow Meow               & 15                                   & Woman               & Asian/Pacific Islander                                                                 & Everyday                                                                          & CS, AI/ML                                                                 \\
Mordecai                & 15                                   & Man                 & Asian/Pacific Islander                                                                 & Once a week                                                                       & CS                                                                        \\
Ricardo                 & 15                                   & Man                 & \begin{tabular}[c]{@{}l@{}}Hispanic/Latinx, \\ Asian/Pacific Islander\end{tabular}     & Everyday                                                                          & CS, AI/ML, AIC                                                            \\
Shark                   & 15                                   & Man                 & Black/African American                                                                 & Every other week                                                                  &                                                                           \\
Tyler                   & 15                                   & Man                 & Black/African American                                                                 & Once a week                                                                       & CS, AIC                                                                   \\ \hline
\multicolumn{6}{l}{\begin{tabular}[c]{@{}l@{}}CS = prior computing education, including workshops and computing courses at school \\ AI/ML = prior AI/ML education, including workshops and AI/ML courses at school \\ AIC = has learned about AI/ML in school (e.g., language arts teacher talked about ChatGPT in one lesson)\end{tabular}}
\end{tabular}
\end{table}

\subsection{Research Approach and Positionality}

We aimed to position high school students as protagonists \cite{iversen2017child}, or main decision-makers, in the process of building LMs. To scaffold such a participatory design sprint within a week \cite{10.1145/3661455.3669876}, we partnered with educators at the local science center. One author has worked with teens at the science center for 10+ years, and another for five. One author is a science center educator who participated in the design, implementation, and analysis. 

We also recognize that our own identities and backgrounds shape our research approach. Our team represents at least five racial/ethnic identities, three gender identities, and academic expertise in the learning sciences, human-computer interaction (HCI), and science education. Our team resides in the same city as the teens.

\subsection{Workshop Activities}  

This paper focuses on the first week of the summer workshop, during which teenagers learned about and designed their own generative LMs. 
Prior to the summer, teens attended a series of four workshops in the spring, in which they reflected on their own use of LMs, learned about Markov chains, and participated in pre-interviews. 
During the summer, each workshop day started with a 30-minute game or warm-up (e.g., duck-duck-goose) and included a snack break. Workshop activities were facilitated by five adults, including an experienced science center educator, two graduate student researchers, and two undergraduate research assistants.
Each day was scheduled as follows (see Figure \ref{fig:activities} for an overview):

\begin{figure}
    \centering
    \includegraphics[width=0.85\linewidth]{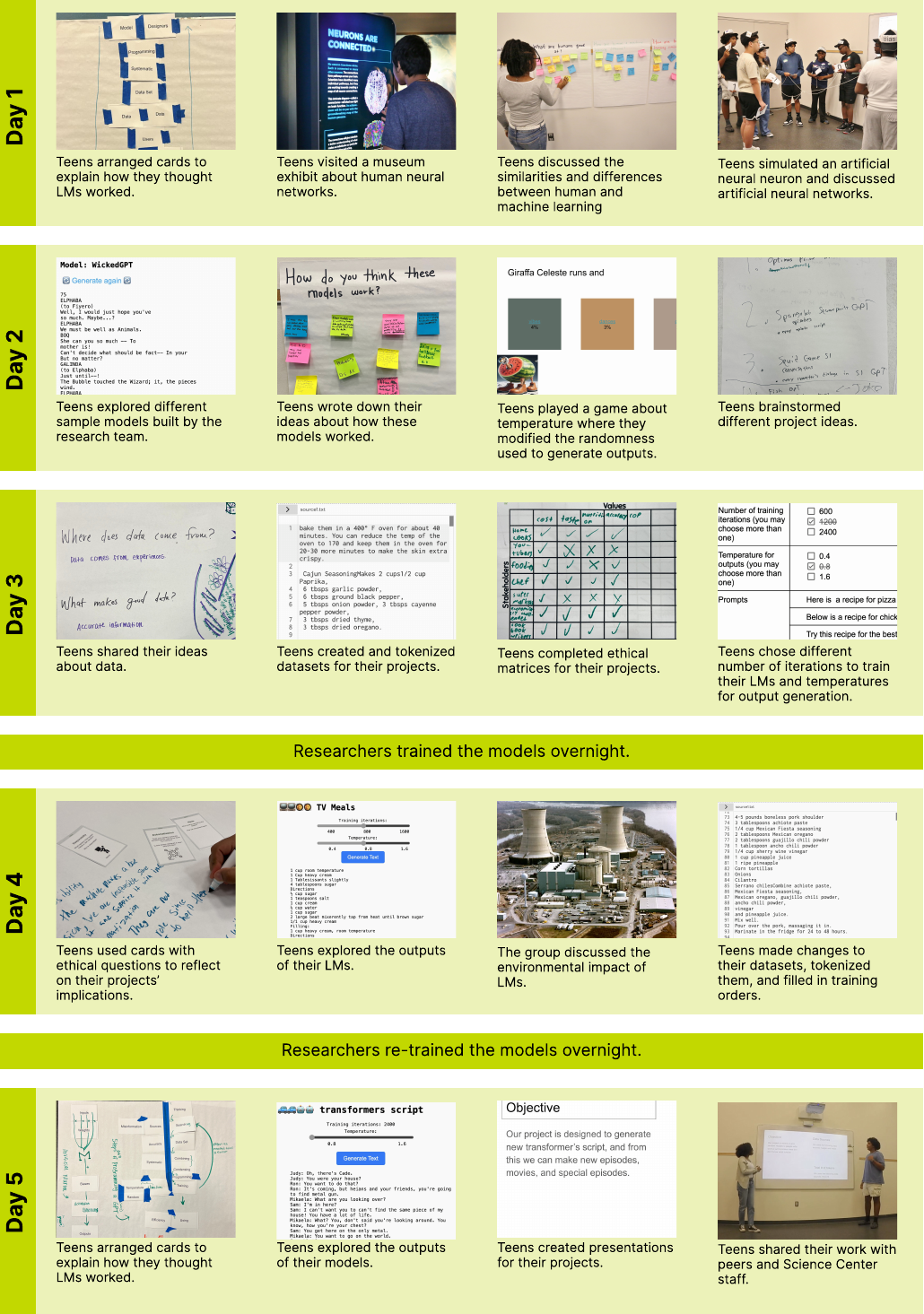}
    \caption{Workshop activities by day. }
    \Description{Figure showing all workshop activities by day. Day 1: Teens arranged cards to explain how they thought LMs worked, Teens visited a museum exhibit about human neural networks, Teens discussed the similarities and differences between human and machine learning, Teens simulated an artificial neural neuron and discussed artificial neural networks. Day 2: Teens explored different sample models built by the research team, Teens wrote down their ideas about how these models worked, Teens played a game about temperature where they modified the randomness used to generate outputs, Teens brainstormed different project ideas. Day 3: Teens shared their ideas about data, Teens created and tokenized datasets for their projects, Teens completed ethical matrices for their projects, Teens chose different number of iterations to train their LMs and temperatures for output generation. Day 4: Teens used cards with ethical questions to reflect on their projects’ implications, Teens explored the outputs of their LMs, The group discussed the environmental impact of LMs, Teens made changes to their datasets, tokenized them, and filled in training orders. Day 5: Teens arranged cards to explain how they thought LMs worked, Teens explored the outputs of their models, Teens created presentations for their projects, Teens shared their work with peers and Science Center staff.}
    \label{fig:activities}
\end{figure}

\paragraph{\textbf{Day 1: Human and artificial neural networks}} Teens completed a card-sorting activity, an activity modeled after a card-sorting task for eliciting folk theories~\cite{10.1145/3173574.3173694}, where they sorted cards with words related to AI/ML.\footnote{Words on the cards drew from vocabulary voiced by the teens in the pre-interviews.}
We asked them to arrange the cards to represent how they thought AI/ML systems worked, using (or not using) as many as they wanted. They were also provided with blank cards to add new words. We explained that there was no right or wrong way to sort the cards and that the purpose was to learn about their thinking.

The team introduced the history of artificial neural networks and their inspiration from human neural networks. 
Following, the teens visited the center's “My Brain” exhibit with the bioscientist who designed it, followed by a discussion about human and machine learning. 
The final activity of the day was an embodied simulation of a single artificial neuron, where different teens played the roles of inputs, weights, bias, a perceptron, an activation function, and outputs.\footnote{The activity was supported by a simple perceptron calculator web application.}

\paragraph{\textbf{Day 2: Temperature and training iterations}} The day began with a review of Markov chains, which the teens had explored in the spring workshops, and a comparison between Markov chains and GPTs. During the discussion, facilitators drew connections to the neural network activities from the previous day, highlighting GPTs as an advanced type of neural network. Following, the teens explored different sample models built by the research team (small LMs trained using the NanoGPT framework).\footnote{The examples included an Obama speech generator, a Friends screenplay generator, a Wicked screenplay generator, and a Presidential speech generator.} The teens were asked to brainstorm how these models worked using the Big Papers method \cite{yip2013brownies}, prompted with the following questions: ``How do you think these models work?'', ``What are they useful for? What worked? Why?'', and ``What are they not useful for? What didn’t work? Why?''

The teens compared outputs generated for nine different LMs trained on the same dataset (screenplays from the TV show Friends~\footnote{Training dataset: 1285903 tokens, validation dataset: 142965 tokens.}), with different training iterations\footnote{A training iteration is a single round of the training process where the model processes one batch of data and updates its weights.} (400, 800, 1600) and different model temperatures\footnote{Temperature is a parameter that controls the randomness of the generated text} (0.4, 0.8, 1.6). After exploring outputs, teens brainstormed how training iterations and temperature influence outputs. Facilitators explained that training iterations enable models to adjust their weights, making connections to the artificial neuron activity from Day 1, and that temperature introduces randomness to text generation. Following, teens generated stories about Italian Brainrot characters\footnote{Italian Brainrot is a set of internet memes that emerged in 2025.} using an interactive tool in which they could control the temperature parameter to observe how temperature influenced the randomness of the output, word-by-word. Finally, teens brainstormed three different ideas for the LMs they wanted to create. 

\paragraph{\textbf{Day 3: Creating datasets to train models}} Day 3 began with another Big Paper activity \cite{yip2013brownies} in which teens shared their ideas about the role of data in LMs. Following, the head of the archives of the science center shared tips about intentionally building datasets. A facilitator introduced \citeauthor{olari2024data}'s \cite{olari2024data} AI/ML data framework, noting the critical role that data plays at all stages of development of LMs. Then, teens read articles about data workers, their role, and the need for humane labor practices. 

The teens returned to their LM ideas and chose one to pursue. Based on their chosen ideas, they created an ethical matrix~\citep{payne2019ethics}.\footnote{Ethical matrices are two-dimensional tables where relevant parties and values are listed along two axes, and designers identify how values align with different parties, noting potential conflicts. This tool was adapted from bioethics to the development of AI/ML systems~\citep{o2017weapons} and later to AI/ML education~\citep{payne2019ethics}.}
Following, the teens created and tokenized\footnote{Tokenization is the process of transforming text into a sequence of tokens, which can be words, characters, or subwords. This is a necessary step in preparing data for training; in tokenizing data, each token is assigned a unique integer used to create a vector representation that is then used in training a model.} a dataset for their own models. Facilitators demonstrated the process, which used a provided JavaScript script using a natural language processing library~\citep{Howe_RiTaJS_2025}. In groups of two to three, the teens were tasked to design datasets for their projects (20,000 to 300,000 tokens). After creating their datasets, the teens filled in ``training orders,'' specifying how many iterations they wanted their models to be trained for (2400, 1200, or 600 iterations), what temperature their models should generate with (0.8, 0.4, or 1.6), and up to four prompts (introduced as ``seeds'' for generating text). Because the computers at the science center did not have enough computing power to train models, the models were trained by three researchers overnight using Kaparthy's~\citep{KaparthyNano} nanoGPT framework\footnote{nanoGPT is a reimplementation of OpenAI's GPT-2 to quickly train and fine-tune medium-sized GPTs.} on M2 and M3 Apple silicon chips (for more details, see \citet{willison_nanogpt_shakespeare_2023}). 

\paragraph{\textbf{Day 4: Iterating on models and reflecting on impact}} To scaffold and make visible the teens' thinking about socio-ethical issues related to LMs, we adapted \citeauthor{bilstrup2020staging}'s \cite{bilstrup2020staging} ethical cards activity. We provided teens with eight cards with questions about data collection, automation, transparency, discrimination, power balance, visibility, and harm. The teens picked three cards and addressed the questions on the cards in relation to their projects. For example, the card titled ``harm'' included the following questions: ``Who could be harmed by this system? How can we prevent harm?'' Each group created and presented a poster discussing potential socio-ethical concerns about their projects. After each presentation, the groups answered questions from their peers. 

Following, the teens explored the outputs of their own models (see Table \ref{tab:projects} for details of each project). On a worksheet, they reflected on how training iterations, temperature, training data, and prompts shaped the outputs. Then, they evaluated each other's projects, addressing the following questions: ``What did you notice in other people’s models?'', ``Which models work better?'', and ``Why do you think these worked better?'' After peer evaluation, they brainstormed how to improve their own models. Researchers facilitated a discussion about the environmental impact of LMs through news articles and a conversation about the energy cost of training their own LMs.\footnote{While training the models overnight, researchers measured the energy consumed using a watt meter and shared with the teens that training their projects took on average 0.05 KWh, approximately the full battery capacity of a laptop for a model trained for 2400 iterations.}
The teens refined their datasets for researchers to retrain their models overnight (see Table \ref{tab:projects} for changes in projects). 

\paragraph{\textbf{Day 5: Final presentations}}
The final day included four main activities: a card sorting activity, an exploration of the retrained models, preparation of project presentations, and presentations of their models.

\subsection{Data Collection and Analysis}

To identify the socio-ethical and technical pieces of knowledge that teens exhibited while participating in the construction of LMs, we took a moment-to-moment microgenetic approach to study team interactions and conversations in the design process. Such an approach allows researchers to ``observe learning processes as they occur and do so in such a way as to permit strong inferences about learning processes'' \cite{chinn2014microgenetic}. 
In CCI, the microgenetic approach has been applied using thematic analysis to study children's understanding of programming \cite{10.1145/2930674.2930725} or scientific concepts when interacting with tangible interfaces \cite{10.1145/1999030.1999044}.  

We collected video recordings of group conversations, screen recordings of teens' work on project computers, and artifacts generated by teens (posters, datasets, models, and presentations). Similar to other studies that have investigated teens’ understandings of AI/ML systems \cite{schaper2023five}, we used thematic analysis to identify teens’ pieces of understanding.
Specifically, we wanted to capture their technical and socio-ethical understandings when brainstorming, creating datasets, evaluating outputs, and discussing socio-ethical issues related to their projects. As such, we focused on the last three days of the workshop.  
To prepare the data for analysis, two researchers created videologs of 26 hours of video and screen recordings. First, the two researchers who videologged the data noted down some possible themes. Second, three researchers went through the data, creating inductive themes (with each theme being a possible piece of understanding). Following, the researchers grouped these themes into pieces of understanding. The coding process was iterative, with several check-ins and discussions with the larger research group. Finally, the three researchers applied the pieces to the data. The goal of the analysis was to identify all the possible pieces exhibited by participants, not their frequency or prevalence; as such, we did not quantify our findings. Overall, we identified ten technical pieces and ten socio-ethical pieces of understanding; these pieces are presented in the findings (see Table \ref{tab:rqs_findings}). 

\section{Findings}

\begin{figure}
    \centering
    \includegraphics[width=1\linewidth]{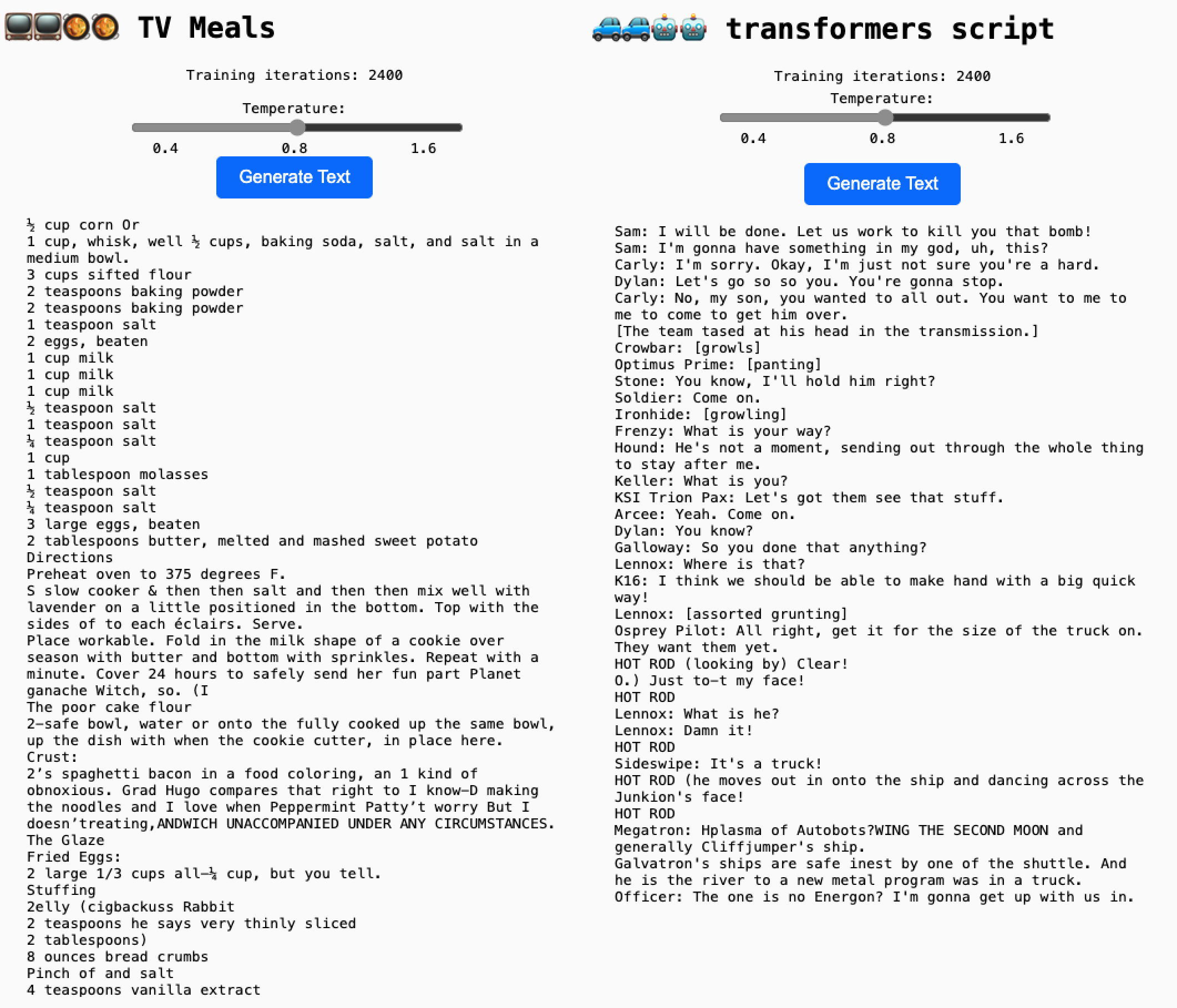}
    \caption{Screenshots of the TV Meals recipe generator designed by Dunkin Lover and Darwin (left) and the Transformer Script generator designed by Meow Meow, Gregory, and Jimmy Bob (right).}
    \Description{Figure showing two outputs of teen-created LMs. TV Meals screenshot: Training iterations: 2400 Temperature: 0.8
    Output
0.5 cup corn Or
1 cup, whisk, well ½ cups, baking soda, salt, and salt in a medium bowl.
3 cups sifted flour
2 teaspoons baking powder
2 teaspoons baking powder
1 teaspoon salt
2 eggs, beaten
1 cup milk
1 cup milk
1 cup milk
teaspoon salt teaspoon salt teaspoon salt
1
1
Cup tablespoon molasses teaspoon salt teaspoon salt
3 large eggs, beaten
2 tablespoons butter, melted and mashed sweet potato
Directions
Preheat oven to 375 degrees F.
S slow cooker & then then salt and then then mix well with lavender on a little positioned
and the betton
bottom owe withe
sides of to each éclairs. Serve.
Place workable. Fold in the milk shape of a cookie over season with butter and bottom with sprinkles. Repeat with a minute. Cover 24 hours to safely send her fun part Planet ganache Witch, so. (I
The poor cake flour
2-safe bowl, water or onto the fully cooked up the same bowl, up the dish with when the cookie cutter, in place here.
2's spaghetti bacon in a food coloring, an 1 kind of obnoxious. Grad Hugo compares that right to I know-D making the noodles and I love when Peppermint Patty't worry But I doesn'treating, ANDWICH UNACCOMPANIED UNDER ANY CIRCUMSTANCES.
The Glaze
Fried Eggs:
2 large 1/3 cups all- cup, but you tell.
Stuffing
Zelly (cigbackuss Rabbit
2 teaspoons he says very thinly sliced
2 tablespoons)
8 ounces bread crumbs
Pinch of and salt
4 teaspoons vanilla extract
Transformers screenshot:
Training iterations: 2400
Temperature:
0.8
Output:
Sam: I will be done. Let us work to kill you that bomb!
Sam: I'm gonna have something in my god, uh, this?
Carly: I'm sorry. Okay, I'm just not sure you're a hard.
Dylan: Let's go so so you. You're gonna stop.
Carly: No, my son, you wanted to all out. You want to me to me to come to get him over.
[The team tased at his head in the transmission.]
Crowbar: [growls]
Optimus Prime: [pantingl
Stone: You know, I'll hold him right?
Soldier: Come on.
Ironhide: [growling]
Frenzy: What is your way?
Hound: He's not a moment, sending out through the whole thing to stay after me.
Keller: What
is you?
KSI Trion Pax: Let's got them see that stuff.
Arcee: Yeah. Come on.
Dylan: You know?
Galloway: So you done that anything?
Lennox: Where is that?
K16: I think we should be able to make hand with a big quick
Lennox: [assorted grunting]
Osprey Pilot: All right, get it for the size of the truck on.
They want them yet.
HOT ROD (looking by) Clear!
0.) Just to-t my face!
HOT ROD
Lennox: What is he?
Lennox: Damn it!
HOT ROD
Sideswipe: It's a truck!
HOT ROD (he moves out in onto the ship and dancing across the Junkion's face!
HOT ROD
Megatron: Hplasma of Autobots?WING THE SECOND MOON and generally Cliffjumper's ship.
Galvatron's ships are safe inest by one of the shuttle. And he is the river to a new metal program was in a truck.
Officer: The one is no Energon? I'm gonna get up with us in.

}
    \label{fig:transformer}
\end{figure}

Throughout the five days of workshop activities, teens iteratively and collaboratively constructed their own LMs, varying in topics and scope (see Table \ref{tab:projects}). Two teams designed models to generate recipes: one used datasets from Kaggle (Recipe Generator), while the other (TV meals) carefully curated recipes related to food depicted on children's TV shows (e.g., spaghetti tacos from iCarly). Two teams created screenplay generators: one on Transformers' movie scripts (see Figure \ref{fig:transformer}), and the other (BlueGPT) mixed screenplays of Bluey\footnote{Bluey is an Australian animated children's show.} with horror movies to create Bluey-themed horror scripts. The GradSpeechGPT team created a graduation speech generator. LyricsGPT designed a pop music lyric generator, and finally, one group created an animal facts generator based on text from National Geographic. The datasets used to train the models varied greatly in size, from 22,790 tokens to 226,484 tokens. 

\begin{table}[]
\small
\caption{Projects, descriptions, and model details}
\label{tab:projects}
\begin{tabular}{@{}llll@{}}
\toprule
\textbf{\begin{tabular}[c]{@{}l@{}}Project\\ (Designers)\end{tabular}} &
  \textbf{\begin{tabular}[c]{@{}l@{}}Project Description \&\\ Changes Made from Day 4 to Day 5\end{tabular}} &
  \textbf{\begin{tabular}[c]{@{}l@{}}Day 4\\ Models\end{tabular}} &
  \textbf{\begin{tabular}[c]{@{}l@{}}Day 5\\ Models\end{tabular}} \\ \midrule
\begin{tabular}[c]{@{}l@{}}TV Meals\\ (Designed by:\\ Dunkin Lover,\\ India \& Darwin)\end{tabular} &
  \begin{tabular}[c]{@{}l@{}}Project: A generator that ``provides consumers\\ with recipes from a variety of tv shows such as\\ ICarly, Teen Titans Go, and Ratatouille''\\ \\ Changes: Organized the recipes better so they\\ were laid out easier to read.\end{tabular} &
  \begin{tabular}[c]{@{}l@{}}Tokens: 22790\\ Iterations:\\ 400, 800, 1600\\ Temperature:\\ 0.4, 0.8, 1.6\end{tabular} &
  \begin{tabular}[c]{@{}l@{}}Tokens: 22794 \\ Iterations:\\ 2400\\ Temperature:\\ 0.4, 0.8, 1.6\end{tabular} \\ \midrule
\begin{tabular}[c]{@{}l@{}}Recipe Generator\\ (Designed by:\\ Butterfly \&\\ Drummer Boy)\end{tabular} &
  \begin{tabular}[c]{@{}l@{}}Project: A generator that ``gives random\\ recipes to people who want a yummylicious\\ meal but don’t know what cuisine''\\ \\ Changes: Added spaces; Deleted gibberish.\end{tabular} &
  \begin{tabular}[c]{@{}l@{}}Tokens: 186516\\ Iterations:\\ 2400\\ Temperature:\\ 0.8\end{tabular} &
  \begin{tabular}[c]{@{}l@{}}Tokens: 186257\\ Iterations:\\ 600, 2400\\ Temperature:\\ 0.8\end{tabular} \\ \midrule
\begin{tabular}[c]{@{}l@{}}Transformers Scripts\\ (Designed by:\\ Meow Meow,\\ Gregory \&\\ Jimmy Bob)\end{tabular} &
  \begin{tabular}[c]{@{}l@{}}Project: A screenplay generator ``that creates\\ Transformers scripts''\\ \\ Changes: Replaced explicit words;\\ Increased training iterations\end{tabular} &
  \begin{tabular}[c]{@{}l@{}}Tokens: 217879\\ Iterations:\\ 2400\\ Temperature:\\ 0.8, 1.6\end{tabular} &
  \begin{tabular}[c]{@{}l@{}}Tokens: 217835\\ Iterations:\\ 2400\\ Temperature:\\ 0.4, 0.8, 1.6\end{tabular} \\ \midrule
\begin{tabular}[c]{@{}l@{}}GradSpeechGPT\\ (Designed by:\\ Briana \& Mordecai)\end{tabular} &
  \begin{tabular}[c]{@{}l@{}}Project: A valedictorian speech generator\\ \\ Changes: Removed anything personal from\\ dataset (names of relatives, years, etc)\end{tabular} &
  \begin{tabular}[c]{@{}l@{}}Tokens: 43745\\ Iterations:\\ 600, 1200, 2400\\ Temperature:\\ 0.4, 0.8, 1.6\end{tabular} &
  \begin{tabular}[c]{@{}l@{}}Tokens: 41000\\ Iterations:\\ 2400\\ Temperature:\\ 0.4, 0.8, 1.6\end{tabular} \\ \midrule
\begin{tabular}[c]{@{}l@{}}BlueGPT\\ (Designed by:\\ Jason \& Tyler)\end{tabular} &
  \begin{tabular}[c]{@{}l@{}}Project: A screenplay generator that\\ ``combines Bluey episodes with horror movies\\ to create a Bluey horror movie or episode''\\ \\ Changes: Removed hyperlinks, explicit\\ language, and one movie; Added 20 Bluey\\ episodes\end{tabular} &
  \begin{tabular}[c]{@{}l@{}}Tokens: 226484\\ Iterations:\\ 2400\\ Temperature:\\ 0.8, 1.6\end{tabular} &
  \begin{tabular}[c]{@{}l@{}}Tokens: 180826\\ Iterations:\\ 600, 2400\\ Temperature:\\ 0.4, 0.8, 1.6\end{tabular} \\ \midrule
\begin{tabular}[c]{@{}l@{}}lyricsGPT\\ (Designed by:\\ King \& Shark)\end{tabular} &
  \begin{tabular}[c]{@{}l@{}}Project: A generator that ``gives people song\\ lyrics, to make the process of writing music\\ easier and more streamlined''\\ \\ Changes: Changed the data sources used\end{tabular} &
  \begin{tabular}[c]{@{}l@{}}Tokens: 93616\\ Iterations:\\ 2400\\ Temperature:\\ 0.8\end{tabular} &
  \begin{tabular}[c]{@{}l@{}}Tokens: 21371\\ Iterations:\\ 600, 2400\\ Temperature:\\ 0.4, 0.8, 1.6\end{tabular} \\ \midrule
\begin{tabular}[c]{@{}l@{}}Animal Facts\\ (Designed by:\\ Ricardo)\end{tabular} &
  \begin{tabular}[c]{@{}l@{}}Project: A generator that ``provides useful\\ information about any kind of animal\\ from the real world''\\ \\ Changes: Removed all references and\\ symbols from dataset\end{tabular} &
  \begin{tabular}[c]{@{}l@{}}Tokens: 85643\\ Iterations:\\ 600, 2400\\ Temperature:\\ 0.8, 1.6\end{tabular} &
  \begin{tabular}[c]{@{}l@{}}Tokens: 53707\\ Iterations:\\ 600, 2400\\ Temperature:\\ 0.4, 0.8, 1.6\end{tabular} \\ \bottomrule
\end{tabular}
\end{table}

In the following sections, we introduce the different technical and socio-ethical pieces of understandings that teens exhibited during the workshop. Following, we highlight how these pieces related to each other. 

\subsection{Technical Understandings In-pieces}
During the workshop, teens exhibited several technical pieces (T) of understandings. While some of these were explicitly discussed in workshop activities (e.g., the artificial neuron simulation), participants expressed these pieces of understandings beyond such activities, using them to explain the functionality of the LMs they constructed. These included pieces about training data, neural networks, and how LMs generate outputs. 

\subsubsection{Training data}
Teens voiced several pieces of technical understandings related to dataset design and data quality. Among these was the idea that \textbf{datasets are designed (T1)}. This piece frequently surfaced when teens interacted with and iterated on their own models. For instance, when describing how the outputs of his team's model were shaped, Gregory lucidly explained, ``you create a dataset, and then train on it''. 

For some teams, this involved making decisions about adding more data, removing data, or carefully editing the datasets. For example, teens voiced that \textbf{data quality shapes LMs performance and outputs (T2)}. This was particularly evident when teens iterated on their datasets. Shark explained that his team redesigned their dataset by ``taking out all the explicit words, replacing them with family-friendly words'' so that the outputs would be appropriate for children. 
Other teens thought about data quality in terms of data diversity. Sometimes, these understandings emerged in relation to the personal experiences and identities of teens. For example, when designing a dataset for a recipe generator, Butterfly searched for puff puff recipes, explaining that she wanted to include Nigerian recipes as her family is Nigerian-American. Drummer Boy, whose family is Dominican, also searched for Dominican recipes and added them to the dataset. Such actions show evidence of teens' understandings of how data quality and diversity shape model performance \textbf{(T2)} and taking action to intentionally design datasets \textbf{(T1)} representative of their lived experiences.
These examples demonstrate how iterating on their datasets supported teens in recognizing that datasets are designed and that data quality decisions affect model behaviors.  

\subsubsection{Neural networks}

Teens expressed varying pieces of technical understandings related to the role of neural networks in LMs. For example, \textbf{LMs are trained using networks of artificial neurons (T3)}. On day 5, when asked by a researcher how LMs worked, Drummer Boy explained, ``in the artificial neuron, you have the inputs times weights plus bias before they go to the activation function.'' He further elaborated, saying that LMs are ``not just shuffling everything together and dishing it out. After learning about the artificial neuron, there's a lot more calculation that goes into it.'' Another piece of technical understandings voiced by teens was that \textbf{LMs learn by iteratively adjusting weights (T4)}. On the last day of the workshop, Dunkin Lover explained: ``The weights can go up and down, so that's kind of what's gonna decide if the output is gonna be negative or positive when it's done doing all the math.'' These pieces demonstrate teens' understanding that LMs are not magical black boxes but rather sophisticated systems that use mathematical operations and weight adjustments to generate outputs.

\subsubsection{Output generation} 
Teens voiced different understandings about how temperature,  probability, condensing, and searching data contribute to the generation of outputs. A common piece of technical understandings voiced by teens was that \textbf{temperature, or randomness, shapes the outputs of LMs (T5)}. Mordecai explained ``the temperature is like how random it is.'' 
Jason explained that his team tried different temperatures and compared outputs: 
\begin{quote}
It's like the temperature helps run the model and output. So, too high of a temperature, the model might run different and, like, be all weird, but too low, it might also not be right. Gotta find the right temperature so the model can run efficiently.
\end{quote}
Through exploration of various outputs, teens recognized temperature as a key parameter that allows designers to have control over the randomness of the outputs generated. 

Despite the specificities of how data, neural networks, and temperature shape model behaviors, teens also voiced more naive pieces of technical understandings about how outputs are generated. Some teens expressed that \textbf{LMs predict the most probable next token (T6).} Meow Meow, for instance, explained that outputs depended on ``some probability of word tokens, because remember the patterns and stuff.'' Other teens argued that \textbf{LMs output the most common answer (T7)} or, as Darwin put it, ``it finds the most efficient answer, most common answers.'' Another piece of technical understandings that also emerged is that \textbf{LMs output the average answer (T8)}. Jimmy Bob elaborated, ``It [the model] goes through iterations, make sure it's good, and then you get the average answer.'' Sometimes teens argued that \textbf{LMs generate outputs by condensing data (T9)}, a piece of technical understandings that can be seen as a precursor to thinking about representations and encoding. Meow Meow explained that the model ``takes your dataset, condenses it'' while Jimmy Bob said that models ``mixed the dataset, it combines it and condenses information.'' Another piece of technical understandings that emerged was that \textbf{LMs generate outputs by searching the training dataset (T10)}. Often, teens would hold two or more of these pieces of technical understandings at the same time. For instance, Meow Meow, who argued that models condense data, also explained that models ``search data and then it gives you the output.'' Similarly, Darwin, who was previously quoted for saying that models give you the most common answer, explained ``the AI goes about finding its answers, so it starts with searching, then it thinks of it, and then it finds the most efficient answer.'' These pieces reveal how teens grappled with both intuitive understandings and more refined understandings in their explanations for how outputs are generated. 

\subsection{Socio-ethical Understandings In-pieces}

During the workshop, teens exhibited varying pieces of socio-ethical (SE) understandings related to responsibility, copyright, creativity, efficiency, misinformation, and discrimination. In the following subsections, we introduce each of these pieces.

\subsubsection{Responsibility} Throughout the workshop, teens voiced divergent and often contradictory understandings of who or what is responsible for the behaviors of LMs, including users, designers, and the systems.  

Sometimes, teens argued that \textbf{users are responsible for the behaviors and potential harms of LMs (SE1)}. A clear articulation of this piece came from India, who explained to her teammates, ``the user would be responsible for any [output], because the system provides what the \textit{consumer} asks.'' For her, users of the system that her team designed were consumers with transactional interactions, where users receive what they ask for. Such a view frames LM-powered systems as neutral tools that can be used for good or bad, depending on the users' intentions. A similar perspective was voiced by 
Brianna and Mordecai, the designers of the graduation speech generator, who explained that users are responsible for ``mix[ing] your own stuff and then changing it to your liking'' (Mordecai). Brianna furthered, ``the user is responsible to not take the speech as it is and instead using it for structure and guidance.'' When asked by a peer how users would know what to do, she explained that the team would add a disclaimer below the output. The idea of adding disclaimers, putting the onus on users, was also voiced by Dunkin Lover, who advocated for a notice that made clear that her team (TV meals) was not liable for any illness caused by the recipes generated by their LM. 

Teens also expressed that \textbf{designers are responsible for the behaviors and potential harms of LMs (SE2).} In addition to his initial claim that users were responsible for ``mixing'' outputs, Mordecai argued that if the GradSpeechGPT ``makes a bad decision, we're responsible because we gave it the information it needs.'' Like Mordecai, other teens also ascribed responsibility to the designers of the datasets used to build models. Tyler, for instance, noted that ``the developers were responsible for, like, if they added anything bad,'' connecting how technical decisions related to dataset design have socio-ethical implications.

A unique piece of socio-ethical understandings voiced only by one participant was that \textbf{LMs are responsible for their behaviors and potential harms (SE3).} Ricardo argued that ``if the AI simply gave out the wrong information for some reason, then it would be on the system itself.'' Yet, he also explained that ``if the output is out to harm somebody, then it would be on us, on our end, because we gave it the information and trained it.'' While Ricardo believed that the harms of misinformation were technical faults of the system, other harms to people would be the responsibility of the designers. 

Throughout these examples, we see how pieces of understandings related to responsibility were context-dependent, sometimes with the same teens, like Ricardo, ascribing the responsibility of different harms to different actors.

\subsubsection{Copyright and the use of data on the internet}

Teens expressed two distinct pieces of understandings in relation to copyright and the use of data on the internet. While they argued that \textbf{creating datasets with data available on the internet may violate copyright laws (SE4)}, they also claimed that \textbf{data on the internet can be used to freely train LMs (SE5)}. 

They explained the potential downsides of using copyrighted data to train models. Drummer Boy, for instance, asked Shark: ``What if you get sued for the other people's lyrics?'' Notably, these concerns were primarily related to avoiding personal legal troubles, rather than broader ethical dilemmas of authorship and intellectual property. A different understanding of copyright and data use was voiced by teens who argued that any data available on the internet can be used to train LMs. For example, Drummer Boy argued that if the data is available on Kaggle, an open-source dataset repository, it is permitted to use it to train a model. To illustrate, we provide an example from Ricardo, who challenged another group (Transformers screenplay generator) with questions about copyright:   
\begin{quote}
    \textbf{Ricardo}: How do you ensure that your GPT does not enter copyright infringement?
    
    \textbf{Jimmy Bob}: Because all the scripts are available online for the public anyway, so why would it be copyright if anybody can go in and get it?
    
    \textbf{Ricardo}: What if you're using the same characters to make a new film?
    
    \textbf{Jimmy Bob}: But all the characters are already in the script online.
    
    \textbf{Gregory}: I forgot to mention we are licensed by Hasbro, so we can do whatever we want.
\end{quote}

In this conversation, we see clear evidence of the two pieces of understanding---that using data from the internet can infringe copyright laws (Ricardo, Gregory) and that data from the internet can be used freely to train generative models without permission (Jimmy Bob)---evident in the interaction. 

Another example came from Logan, who argued that data from the internet cannot simply be appropriated for model training because of copyright restrictions \textbf{(SE4)} and because ``we need permission.'' Logan, grounding his understanding of copyright in his lived experiences, explained that he had been involved in the City Community Youth Court, an out-of-school restorative justice program, where he learned that asking for permission to use data is important. Such pieces of understandings about data and copyright were often exhibited when teens interacted with projects their peers created.

\subsubsection{Creativity and ``creative jobs''}

Teens also voiced socio-ethical understandings related to creativity and creative jobs. A frequently surfaced piece was that \textbf{LMs devalue human creativity (SE6)}.
For example, teens considered the effects of LMs at a societal level, highlighting how ``AI slop''\footnote{AI slop is cheap AI/ML-generated content with little intentionality, uncritical deployment, and ``soullessness'' \cite{silbey2025ai}.} may devalue the arts. Brianna explained that using screenplay generators ``could decrease the value of movies in general because they're coming out at a much faster pace.'' Gregory, one of the creators of the screenplay generator, explained that they were concerned the most about ``the replacement of, well, the removal of creativity,'' explaining that, even though they generated a bunch of scripts, these did not seem as creative as human-generated ones. Similarly, Jimmy Bob was particularly concerned with novelty, positing that screenplay generators would only replicate the style and tropes of old movies, as they were trained on past screenplays.

Another piece exhibited by teens was that \textbf{creative jobs would become less valuable due to LMs (SE7)}.\footnote{We acknowledge that all labor is creative \cite{rose2005mind}. We use the terms ``creative jobs'' and ``creative workers'' since these were used by teens to refer to screenplay writers and lyricists.} Shark, one of the creators of the lyricsGPT, acknowledged that ``lyricists will still have the job of, like, checking it, proofreading it, and then, sort of, editing it to fit the artist's own style [...] but it won't be as big a job.'' Similarly, Gregory openly accepted that their screenplay generator `` will be putting the screenwriters or whatever out of a job, but they will still have the job of correcting these scripts.'' 
Jimmy Boy acknowledged that creative jobs could be less valuable but proposed an alternative, explaining that one way to prevent AI/ML systems from taking away jobs is ``only giving those scriptwriters access to that AI or only allowing scriptwriters to go in and edit it to make it more human-like.'' Here, we see teens understand how generative LMs may transform and restructure creative roles. 

\subsubsection{Efficiency}
A pervasive piece of socio-ethical understandings that mediated teens' understandings of AI/ML was that \textbf{LMs make human tasks more efficient (SE8)}. This idea was brought up by teens across all groups. According to the teens, their LMs could write songs and screenplays more efficiently, make it easier to find information and cook food, or create graduation speeches more quickly. For example, even Jimmy Bob, who argued for workers' control of LMs, noted that using a generative model could ``shave a couple of weeks'' of work off by giving writers a ``baseline of what the movie plot could be.'' 
Ricardo also emphasized efficiency as the justification for his animal fact generator, claiming that instead of reading many articles written by expert users, people who want to learn about animals could ``learn quickly about it'' through short-form text generated by their LM.

\subsubsection{Misinformation and discrimination}
Teens also surfaced common perceptions that misinformation and discrimination are undesirable in LMs. The \textbf{misinformation is undesirable (SE9)} piece surfaced when teens talked about creating datasets. Meow Meow, for instance, acknowledged that dataset designers have the responsibility ``to look out for misinformation and the accuracy of the information you were feeding it.'' The \textbf{discrimination is undesirable (SE10)} piece was voiced when considering the possible implications of LMs: ``the lyric generator can cause discrimination, which can harm people'' (Jason).
These pieces show that learners’ understandings of LMs included prioritizing the mitigation of misinformation and discrimination as essential components of model design.

\subsection{Connecting Pieces and Building Understanding}

\begin{figure}
    \centering
    \includegraphics[width=1\linewidth]{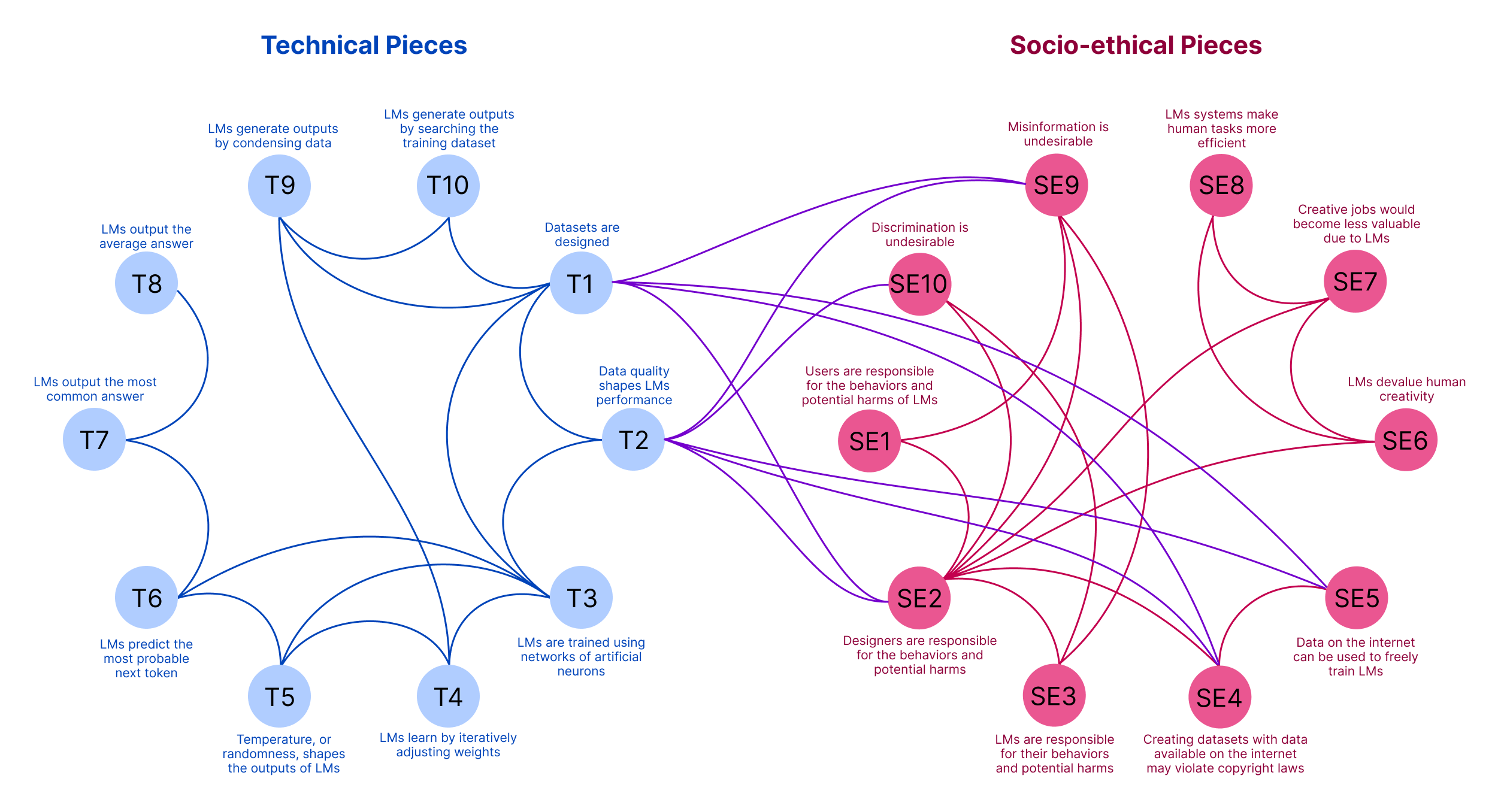}
    \caption{Connections between technical pieces (blue edges), connections between socio-ethical pieces (red edges), and connections between a technical and a socio-ethical pieces (purple edges).}
    \Description{Image shows technical pieces in blue and socio-ethical pieces in red with lines connecting those pieces that relate to each other. For an accessible version see Table \ref{tab:rqs_findings}}
    \label{fig:network}
\end{figure}

Building LMs provided teens with a context to integrate and connect pieces together to build understandings of how AI/ML systems work. In this section, we provide evidence of how socio-ethical and technical pieces related to each other and highlight contradictions that emerged in teens' understandings of LMs. Throughout the workshop, we observed learners make connections between pieces of knowledge, building their understandings of LMs while constructing their own models. While we do not have the space to elaborate on every connection, Table \ref{tab:rqs_findings} and Figure \ref{fig:network} show the different relationships between pieces of knowledge that we identified. In the following paragraphs we describe some of these connections. 

\begin{table}[htbp]
\centering
\small
\caption{Pieces and their relationships to each other}
\label{tab:rqs_findings}
\begin{tabular}{l p{7.75cm} p{2.75cm} p{2.5cm}}
\toprule
\textbf{Code} & \textbf{Pieces of Understanding} & \textbf{Related Socio-ethical Pieces} & \textbf{Related Technical Pieces} \\
\midrule
T1 & Datasets are designed & SE2, SE4, SE5, SE9, SE10 & T2, T3, T9, T10 \\

T2 & Data quality shapes LMs performance and outputs & SE2, SE4, SE5, SE9, SE10 & T1, T3 \\

T3 & LMs are trained using networks of artificial neurons &  & T1, T2, T4, T5, T6 \\

T4 & LMs learn by iteratively adjusting weights &  & T3, T5, T9 \\

T5 & Temperature, or randomness, shapes the outputs of LMs &  & T3, T4, T6 \\

T6 & LMs predict the most probable next token &  & T3, T5, T7 \\

T7 & LMs output the most common answer &  & T6, T8 \\

T8 & LMs output the average answer &  & T7 \\

T9 & LMs generate outputs by condensing data &  & T1, T4, T10 \\

T10 & LMs generate outputs by searching the training dataset &  & T1, T9 \\

SE1 & Users are responsible for the behaviors and potential harms of LMs &  SE2, SE9 &  \\

SE2 & Designers are responsible for the behaviors and potential harms of LMs systems & SE1, SE3, SE4, SE6, SE7, SE9, SE10 & T1, T2 \\

SE3 & LMs are responsible for their behaviors and potential harms & SE2, SE9, SE10 &  \\

SE4 & Creating datasets with data available on the internet may violate copyright laws & SE2, SE5 & T1, T2 \\

SE5 & Data on the internet can be used to freely train LMs & SE4 & T1, T2 \\

SE6 & LMs systems devalue human creativity & SE2, SE7, SE8 &  \\

SE7 & Creative jobs would become less valuable due to LMs & SE2, SE6, SE8 &  \\

SE8 & LMs systems make human tasks more efficient & SE6, SE7 &  \\

SE9 & Misinformation is undesirable & SE1, SE2, SE3 & T1, T2 \\

SE10 & Discrimination is undesirable & SE2, SE3 & T2 \\

\bottomrule
\end{tabular}
\end{table}

\begin{figure}
    \centering
    \includegraphics[width=1\linewidth]{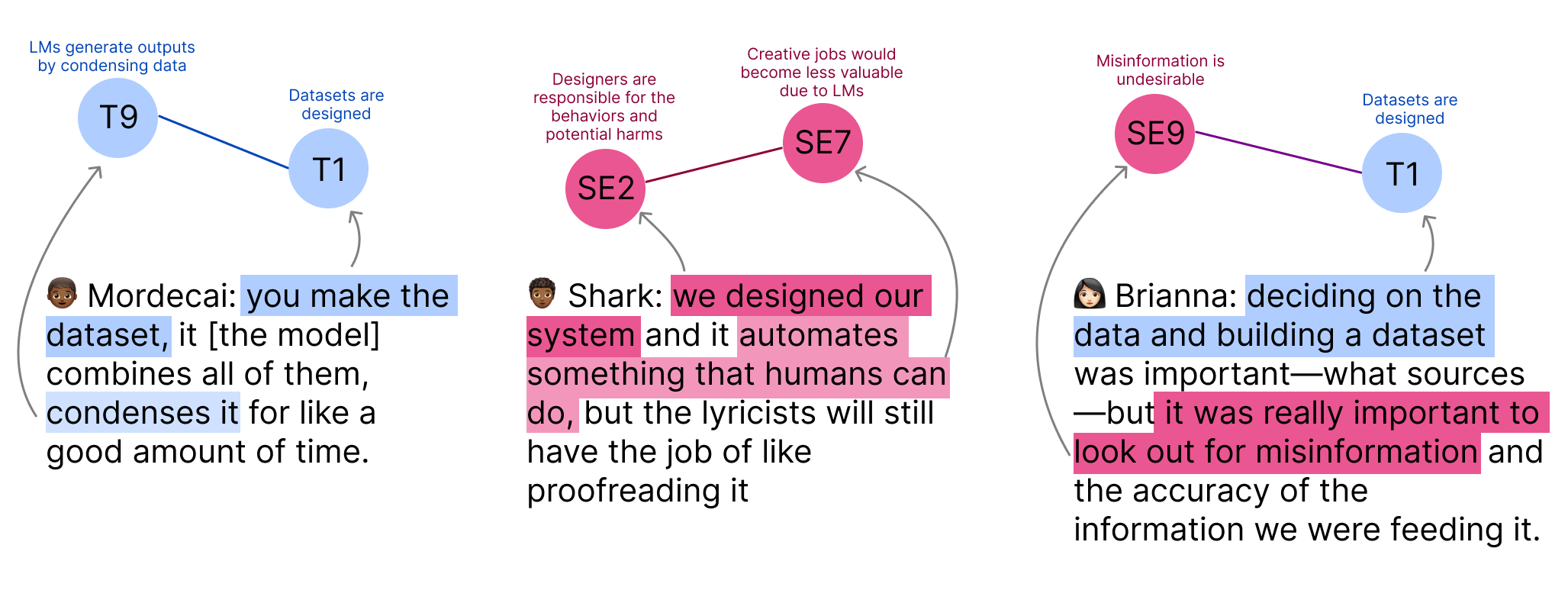}
    \caption{Examples of connected technical pieces (left), connected socio-ethical pieces (center), and the connection between a technical and a socio-ethical pieces (right).}
    \Description{
    three-column infographic.
    Column 1: Quote from Mordecai ``Mordecai: you make the dataset, it [the model] combines all of them, condenses it for like a good amount of time.'' Quote is connected to circles labeled T9 and T1. 
    Columm 2: Quote from Shark: ``we designed our system and it automates something that humans can do, but the lyricists will still have the job of like proofreading it.'' Quote is connected to circles SE2 and SE7
Column 3: Quoute from Brianna: ``deciding on the data and building a dataset was important—what sources—but it was really important to look out for misinformation and the accuracy of the information we were feeding it.'' Quote is connected to circles SE9 and T1.    
    }
    \label{fig:connections}
\end{figure}

Such connections between technical pieces of knowledge were particularly evident among pieces related to datasets \textbf{(T1, T2)}. For instance, the idea that datasets are designed \textbf{(T1)} was often expressed in relation to how data quality shapes performance and outputs \textbf{(T2)}. Gregory and Tyler both voiced their annoyance at the process of designing a dataset because they had to ensure that the data was of good quality. ``That was just tedious!'' Gregory complained when cleaning the data and removing inappropriate words. Similarly, Tyler explained that ``filtering all the bad words we had in our scripts was probably the hardest thing that we did.'' Such comments make clear the connection between their understanding of datasets being designed and how data quality shapes model behaviors. Often, the idea that datasets are designed \textbf{(T1)} was also related to teens' understandings that LMs condense data \textbf{(T9)} or search the training dataset \textbf{(T10)}. We can see an example of the relationship between \textbf{T1} and \textbf{T9} when Mordecai explained: ``you make the dataset, it [the model] combines all of them, condenses it for like a good amount of time'' (see Figure \ref{fig:connections}).

Socio-ethical pieces were also voiced in relation to each other. This was particularly evident when teens ascribed responsibility for LM behaviors and potential harms to designers \textbf{(SE2)}. This piece was often voiced in relation to potential copyright violations \textbf{(SE4)}, devaluing human creativity and creative jobs \textbf{(SE6, SE7)}, and misinformation and discrimination being undesirable \textbf{(SE9, SE10)}. For example, consider the following reflection by Shark: ``we designed our system, and it automates something that humans can do, but the lyricists will still have the job of like checking it, proofreading it'' (see Figure \ref{fig:connections}). Here, voiced that designers are responsible for the system \textbf{(SE2)} in relation to the automation and transformation of creative jobs \textbf{(SE7)}. Another example of \textbf{(SE2)} being related to another socio-ethical piece came from Gregory, who voiced: ``we are licensed by Hasbro, so we can do whatever we want.'' This sentence shows Gregory's understanding that designers are responsible \textbf{(SE2)} and that using data from the internet without the required permissions could lead to copyright violations \textbf{(SE4)}. 

Technical and socio-ethical pieces were also often connected to each other as learners built understandings about the models they created. This was especially clear when teens explained how the design of datasets \textbf{(T1)} related to socio-ethical issues, including designers being responsible for system behaviors \textbf{(SE2)}, possible copyright violations \textbf{(SE4)},  using data from the internet \textbf{(SE5)}, and misinformation \textbf{(SE9)}. Mordecai, for instance, explained ``if it [the model] makes a bad decision, we're responsible because we gave it, like, the data or information it needs and we supplied it with [them].'' Here, his understanding of the responsibility of designers \textbf{(SE2)} is intertwined with his understanding that datasets are designed \textbf{(T1)}. Another example that illustrates how \textbf{T1} is related to a socio-ethical piece came from Brianna, who explained, ``it's like deciding on the data and building a dataset was important but it was really important to look out for misinformation and the accuracy of the information we were feeding it'' (see Figure \ref{fig:connections}). Here we can see how \textbf{T1} was related to misinformation being undesirable \textbf{(SE9)}.   

The relationships between pieces often also illustrated contradictions in how teens understand LMs. Teens commonly voiced contradictory socio-ethical pieces when talking about their own models and peers' models. For example, Drummer Boy asked Shark, ``What if you get sued for the other people's lyrics?'' This question shows his understanding that using other people's work may be a copyright violation \textbf{(SE4)}. Yet, a few minutes later, he explained: ``we used recipes from the internet; anyone can find that anyone can use it; we just made it more accessible for everyone, easier, no copyrights and nothing like that,’’ when talking about his own model \textbf{(SE5)}. Such contradictions were also present in teens' technical understandings. For instance, consider Meow Meow's explanation of how LMs works: ``It searches data. And then it gives you a dataset, condenses it. And then, once it got all this information, you train it.'' Here, different ideas about how LMs generate outputs are muddled together \textbf{(T9, T10, T3)}. Such contradictions are evidence of how unstable novices' understandings may be and how such understandings, as in the case of Shark, depend on context.

These examples show us that teens' understandings of LMs were fragmented, context-dependent, and sometimes contradictory. At the same time, they illustrate how pieces of technical and socio-ethical understandings are not disjointed; instead, these relate to each other and exist in an ecology of pieces.   
\section{Discussion}

While prior research demonstrates the feasibility of engaging teens in the construction of generative LMs \cite{MORALESNAVARRO2025100769, 10.1145/3713043.3728857, 10.1145/3737609.3747112, 10.1145/3769994.3770061}, such work does not address how construction may support teens in understanding LMs. Through this paper, we contribute an inventory of technical and socio-ethical pieces of understandings of LMs that teens exhibited when participating in LM construction activities. We discuss these understandings in relation to prior research on AI/ML and sensemaking. Following, we argue why an in-pieces approach may be particularly helpful in studying the technical and socio-ethical sensemaking of LMs. Finally, we discuss implications for the design of learning tools and activities.  

\subsection{Teens' Technical and Socio-ethical Understandings}
Research on construction in CCI argues that, when young people build computing systems, they can better understand how these work and the societal implications of such systems \cite{dindler2020computational, iversen2017child, iivari2023computational}. However, there is no prior evidence of the kinds of understandings teens may exhibit when building LMs. We documented a collection of pieces of understandings that teens demonstrated when designing their own LMs, providing evidence of what young people think about LMs while engaging in construction activities. Here, we discuss how our findings relate to prior research on sensemaking of AI/ML systems. 

The technical pieces of understandings exhibited by youth varied from naive elements to more sophisticated ideas about how data, training iterations, and neural networks shape the outputs of LMs. Some pieces of understandings observed in our study had already been examined in prior research looking at young people's understandings when using AI/ML systems. For example, like \citet{marx2024identifying}, we observed teens' understandings that models generate outputs by searching the training dataset \textbf{(T10)}. Additionally, similar to prior research, the teens exhibited understandings that models make predictions \textbf{(T6)} \cite{10.1145/3713043.3728856} and data shapes model performance \textbf{(T2)} \cite{lin2025children}. Our work supports that such pieces of understandings may be common among novice teens and presents a selection of more sophisticated pieces of understandings that have not been documented in prior literature. More importantly, our work emphasizes that naive and more sophisticated pieces coexist in teens' understandings of LMs.

The socio-ethical pieces observed in the study were also diverse. As documented in prior literature, teens in our study also exhibited understandings that discrimination is undesirable \textbf{(SE10)} \cite{10.1145/3762807, heeg2025young,lee2022eliciting,salac2023scaffolding}, misinformation is undesirable \textbf{(SE9)} \cite{10.1145/3762807}, and AI/ML systems may devalue human agency and creativity \textbf{(SE6)} \cite{solyst2024children}. While our work affirms that teens may hold such pieces of understanding, it also highlights the complexity of teens' socio-ethical thinking, which were diverse and divergent. For example, teens ascribed responsibility for the behaviors and potential harms of LMs to varying parties, voiced contradictory understandings about the use of data and copyright, and diverged in their optimism for such technologies.

With such diverse technical and socio-ethical understandings of LMs, we argue, it is necessary to take an ecological perspective to study how these relate to each other and emerge in context.

\subsection{Understandings in Pieces}
In this paper, we took an in-pieces approach that builds on research on knowledge-in-pieces \cite{disessa2004coherence}, ideology-in-pieces \cite{philip2011ideology}, and algorithmic folk theories \cite{10.1145/3173574.3173694}. These theoretical frameworks argue that understanding of phenomena is fragmented, characterized by instability (and sometimes incoherence), and that different pieces of understandings are activated in different contexts. By bringing together work on fragmented understandings of scientific phenomena \cite{disessa2004coherence}, fragmented socio-ethical and ideological thinking \cite{philip2011ideology}, and fragmented understandings of socio-technical systems \cite{10.1145/3173574.3173694} we argue that studying teens' technical and socio-ethical understandings of LMs through an in-pieces approach can be a fruitful approach to unpack how such understandings are built, relate to each other, and emerge in context.

In our study, teens exhibited an extensive collection of pieces depicting novices' unstable and fragmented understandings of LMs. In the context of building their own LMs, these pieces came together in different arrangements that combined various technical and socio-ethical pieces. In the case of technical pieces, these arrangements often included naive pieces (e.g., LMs generate outputs by condensing data \textbf{(T9)} or searching the dataset \textbf{(T10)}) combined with more formal or normative pieces (e.g., LMs are trained using networks of artificial neurons \textbf{(T3)}). While we could dismiss naive pieces as misconceptions and label them as obstacles for learning, like \citet{smith1994misconceptions} who argue that all knowledge is productive for learning, we see value in such thinking as resources for growth, refinement, and reorganization. For example, the idea that LMs output the most common answer \textbf{(T8)} has the potential to be a precursor of other pieces that account for the roles of probability \textbf{(T6)} and temperature \textbf{(T6)} in output generation. The apparent contradictions are evidence of the instability of novice understandings and how, in the context of building models, novices can refine and reorganize their understandings to incorporate more normative and technically accepted explanations.

In the case of socio-ethical pieces, contradictory pieces are evidence that socio-ethical understandings are also unstable and highly context-dependent. For instance, teens found it quite easy to reflect on potential limitations of LMs (e.g., \textbf{SE4, SE6}) when talking about their peers' projects, but oftentimes adopted techno-optimist stances, prioritizing efficiency \textbf{(SE8)}, in their own projects. Here, like \citet{philip2011ideology}, we argue that ideology played a big role in shaping teens' socio-ethical understanding. While we saw fleeting moments of `critical awakening' \cite{malik2022critical} with teens voicing deep ethical concerns about data use \textbf{(SE4)} or imagining the possibility of workers controlling LMs and their use \textbf{(SE7)}, these pieces of understandings were accompanied by pieces rooted in the mainstream techno-optimist ideology driving AI/ML development in the US (e.g., \textbf{SE1, SE5, SE8}). It is worth noting that most socio-ethical pieces were exhibited when participating in activities intentionally designed to elicit reflection---such scaffoldings were crucial in supporting socio-ethical thinking.

This study was a first attempt at studying technical and socio-ethical understandings together using an in-pieces framework. We argue that if CCI and computing education scholars consider technical and socio-ethical aspects of AI/ML as equally important in learning experiences, it is necessary to analyze technical and socio-ethical understandings side-by-side, using a framework that accounts for the interrelated nature of technical and socio-ethical issues in computing. Future research should explore how teens' understandings change over time, mapping changes in how young people build, refine, and connect pieces. Other studies could take a more traditional approach by comparing teens' pieces before and after participating in AI/ML literacy interventions or comparing how different learning activities may support the refinement and connection of different pieces.   

\subsection{Implications for the Design of Learning Tools and Activities}

In our work, iteratively designing datasets, training models for different numbers of iterations, and generating outputs with different temperatures were crucial in supporting learners in making inferences about LMs and how they work. Comparing outputs was particularly productive for learners when observing differences in model behaviors. It was through this iterative process that learners voiced clear pieces of understandings that data content and quality \textbf{(T2)}, training iterations \textbf{(T4)}, and model temperature \textbf{(T5)} shape model behaviors. Although iterative design has been discussed before in CCI research on construction activities \cite{iversen2017child}, AI/ML tools rarely support comparing models or evaluating models. Tools designed to train generative language models should enable learners to iteratively design datasets, play with different model parameters, and systematically compare outputs. Similarly, learning activities should emphasize the importance of iteratively designing and assessing LMs' performance. It is worth noting that designing LMs in this study involved significant participation from researchers who trained the models overnight. Here, we see the need for tools that are accessible and manageable for novice model designers. We are excited for the potential of new interfaces that may support model training and fine-tuning on the browser \cite{10.1145/3769994.3770061}.

In this study, most socio-ethical pieces were exhibited during activities with the explicit goal of reflecting on the limitations, implications, and decisions made in the process of building LMs. Using ethical matrices \cite{payne2019ethics} and ethics cards \cite{bilstrup2020staging} was helpful in scaffolding reflection. As such, we recommend that model design workshops also include time dedicated to thinking about socio-ethical issues. We also argue for the design of tools that can support learners in auditing and empirically investigating the behaviors and limitations of their own LMs. Adapting methods and toolkits that have been used with experts to scaffold socio-ethical thinking \cite{deng2025supporting} may also be productive in supporting youth. 

In relation to both technical and socio-ethical aspects, from a pedagogical perspective, we made decisions to glassbox (data design, neural networks, temperature, copyright and data) and blackbox (embeddings, environmental concerns) certain aspects of LMs. Future research is necessary to assess what aspects of LMs should be made transparent to provide productive contexts for youth to better understand AI/ML systems.

\subsection{Limitations}
We acknowledge that our study of teens' understandings is based on the ideas they expressed by talking or writing during the workshop. It is possible that some teens who did not talk often had different understandings than those who verbalized their ideas. As we have mentioned throughout the paper, understanding is dependent on context; as such, our findings should be interpreted within the context of the workshop and are not intended to be generalized to the broader population. Readers may find our insights applicable to similar contexts. We conducted these activities in the context of a summer program with motivated teens interested in science and technology, but the idea of engaging teenagers in designing LMs could be more impactful if designed for classroom settings with a wider range of learners, not just those with interest and access to an out-of-school program. It is important to replicate these kinds of learning activities and analysis in different contexts in which teens design LMs to gain more evidence about the kinds of understandings they may have.  
\section{Conclusion}

CCI research argues that engaging learners in construction activities supports the development of their understanding of computing systems. Yet, in the case of generative language models, prior research shows the feasibility of engaging teens in designing models without sufficient evidence of the kinds of understandings they may have. We conducted a participatory design workshop in which sixteen teens designed generative language models (e.g., screenplay, recipe, and lyric generators) to investigate their technical and socio-ethical understandings. We documented an inventory of "pieces of understandings" that demonstrate that the teens in the study held a collection of complex, divergent, naive, and sophisticated ideas. Our findings suggest that taking an in-pieces approach is not only a viable one, but also a necessary one, which can consider and analyze technical and socio-ethical understandings in tandem. We hope that this work motivates future research that supports teens in designing generative language models to develop technical and socio-ethical understandings of these systems. Such understandings are necessary for young people to understand the world that they live in and to see AI/ML technologies not as inevitable, but as systems that are designed and maintained, opening doors for them to see themselves as potential designers of these ubiquitous systems.

\section{Selection and Participation of Children}
We recruited teens enrolled in a STEM afterschool program in a city located in the Northeastern United States. Teens were invited by the organizer of the STEM program to participate via email and through paper handouts. Parents received consent forms prior to the study, which included a brief explanation of the research, and the teens assented to their participation. Furthermore, the lead researcher met with parents to answer any questions they had about the workshops and study. Research protocols and data collection methods were approved by the IRB board of the University of Pennsylvania.

\begin{acks}
With regards to Elo Esalomi for their support in data collection. This work was partially supported by National Science Foundation grants (\#2414590, \#2333469), a Penn AI fellowship, and a rapid response grant from the Spencer Foundation, the Kapor Center, the William T. Grant Foundation, and the Alfred P. Sloan Foundation.
\end{acks}

\bibliographystyle{ACM-Reference-Format}
\bibliography{4_references}
\pagebreak 
\end{document}